\documentstyle{article}
\newif\ifdraft
\draftfalse  
\def\note[#1]#2{\message{(#1)}\ifdraft{\noindent\em #2}\fi}
\ifdraft
\fi
\font\diagramvii=diagram5
\font\diagramv=diagram5
\newfam\diagramfam

  \textfont\diagramfam=\diagramvii
  \scriptfont\diagramfam=\diagramvii
  \scriptscriptfont\diagramfam=\diagramv
\def\diagchar#1{\count0="70 \advance\count0by\diagramfam %
  \multiply\count0by"100 \advance\count0by#1 \mathchar\the\count0}
\newcount\diagch
\def\diagop#1from#2to#3.{\diagch=#2 \diagchar\diagch%
  \loop\ifnum\diagch<#3 \advance\diagch by1 #1\mskip0mu plus2mu %
  \allowbreak\diagchar\diagch\repeat}
\def\G{{\cal G}}		    
\def\D{{\Gamma}}		    
\def\H{{\cal H}}		    
\def\Th{\Theta}			    
\def\S{{\cal S}}		    
\def\W{{\cal W}}		    
\def\L{\ell}			    
\def\V{{\cal V}}		    
\def\const{c\cdot}		    
\def\K{{\rm K}}			    
\def\T{{\rm T}}			    
\def\Rop{{\rm R}}		    
\def\Rbar{{\rm\bar R}}		    
\def\deg{\mathop{\rm deg}}	    
\def\dim{\mathop{\rm dim}}	    
\def\implies{\Rightarrow}	    
\def\ddJ{{\delta\over\delta J}}	    
\def\rational#1#2{{\mathchoice{\textstyle{#1\over#2}}%
  {\scriptstyle{#1\over#2}}{\scriptscriptstyle{#1\over#2}}{#1/#2}}}
\def\half{\rational12}			    
\begin{document}

\rightline{FSU--SCRI--96--132}
\vspace{15mm}

\begin{center}
  {\Large\bf A Simple Proof of the BPH Theorem\footnote{To be published in the
    proceedings of QFTHEP '96, St. Petersburg, Russia.}} \\
  \vspace{4mm}
  A. D. Kennedy \\
  Supercomputer Computations Research Institute \\
  Florida State University, Tallahassee, Florida 32306--4052, U.S.A.
\end{center}

\begin{abstract}
  \noindent A new formalism is given for the renormalization of quantum field
  theories to all orders of perturbation theory, in which there are manifestly
  no overlapping divergences. We prove the BPH theorem in this formalism, and
  show how the local subtractions add up to counterterms in the action.
  Applications include the renormalization of lattice perturbation theory, the
  decoupling theorem, Zimmermann oversubtraction, the renormalization of
  operator insertions, and the operator product expansion.
  {\parfillskip=0pt\par}
\end{abstract}

\section{Introduction}

The first approach to showing that the divergences of quantum field theories may
be absorbed into local counterterms to all orders in perturbation theory was due
to Dyson \cite{dyson49a,dyson49b}. The $\Rop$-operation was introduced by
Bogoliubov and Parasiuk \cite{bogoliubov56a,bogoliubov57a} following some
earlier work of St\"uckelberg and Green \cite{stuckelberg51a}, and their proof
that this removed all the divergences was corrected by Hepp \cite{hepp66a}.

Our approach is based on several ideas. The idea of differentiation with respect
to external momenta comes from Tarasov and Vladimirov
\cite{vladimirov80a,tarasov80a} and Chetyrkin, Kataev, and Tkachev
\cite{chetyrkin80a}. The approach to proving the equivalence of the subtractions
made by the $\Rop$-operation and counterterms is due to Anikin, Polivanov, and
Zavialov \cite{anikin73a}. These ideas were combined to provide a proof of the
BPH theorem by Caswell and Kennedy \cite{kennedy82b}, which however did not
provide a completely satisfactory proof that subtracted integrals which were
overall (power-counting) convergent were actually convergent. The idea (but not
the name) of a ``small momentum cutoff'' was introduced by Hahn and Zimmermann
\cite{zimmermann68a}, and the Henge decomposition was introduced by Caswell and
Kennedy \cite{kennedy83a} and applied to the closely related problem of studying
the asymptotic large-momentum behaviour of convergent Feynman diagrams.

The present work combines and extends these methods to give a proof of the BPH
theorem which makes no use of Feynman parameters. This is important as the proof
is applicable to lattice perturbation theory where the propagators are not
quadratic forms in the momenta, and the usual Feynman parameterization is not
applicable \cite{reisz88c,reisz88e}. The present proof is only directly
applicable to theories in Euclidean space: for lattice field theory this is all
that is needed, and for theories with quadratic propagators the corresponding
Minkowski space results follow from the Euclidean space ones
\cite{zimmermann68b}. We also require that there are no massless propagators in
order to avoid infrared divergences.

\section{Graphs and Integrals}

We need to make a few elementary definitions if for no other reason than to
specify our notation. A graph is {\em connected\/} if it cannot be partitioned
into two sets of vertices which are not connected by an edge. A graph is {\em
one particle irreducible\/} (1PI) if it remains connected after removing any
edge. A single vertex is thus a 1PI graph. A {\em Feynman integral\/} $I(\G)$
may be associated with any graph $\G$ by means of the {\em Feynman rules\/} for
the theory. A propagator is associated with each line, some factor with each
vertex, and a $D$-dimensional momentum integral with each independent closed
loop. $I(\G)$ is a function of the external momenta $p$, the lightest mass $m$
(we assume $m>0$ to avoid infrared divergences), some dimensionless couplings,
and a cutoff $\Lambda\equiv1/a$ which is introduced to make the theory well
defined. We extend the mapping $I:\G\mapsto I(\G)$ to act linearly on sums of
graphs.

\subsection{Derivatives of Graphs and Taylor's Theorem}

It is useful to consider the {\em derivative\/} of a Feynman diagram with
respect to its external momenta. This is drawn diagrammatically as
$\partial\diagchar0 = \diagchar{'106} + \diagop+from'107to'116.$. Note that we
view crossed lines and vertices as associated with new Feynman rules: although
one might view the cross as a new vertex inserted into a line this notation is
not adequate in general when vertices (including such crosses themselves) have a
non-trivial momentum dependence.

Each of the graphs shown above is really a sum over all the components of all
the independent external momenta, $I(\partial\G) = {\partial I(\G)\over\partial
p_\mu}$, $I(\partial^2\G) = {\partial^2 I(\G)\over\partial p_\mu\partial
p_\nu}$, and so forth. Viewing $I(\G)$ as a function of its external momenta,
repeated application of the fundamental theorem of calculus gives us Taylor's
theorem. In our notation $I(p) = \T^n I(p) + \int_{p_0}^p dp_1 \int_{p_0}^{p_1}
dp_2 \ldots \int_{p_0}^{p_{n-1}} dp_n \,\partial^n I(p_n)$, where
\begin{displaymath}
  \T^n I(p) \equiv \sum_{j=0}^n {(p-p_0)^j\over j!}\partial^j I(p_0)
    = \sum_{j=0}^n \sum_{\mu_1,\ldots,\mu_j}
      {(p-p_0)_{\mu_1}\ldots (p-p_0)_{\mu_j}\over j!}
      {\partial^j I(p_0)\over \partial p_{\mu_1}\ldots\partial p_{\mu_j}}.
\end{displaymath}

\subsection{Henges}

Any graph may be decomposed into a set of disjoint 1PI components and a set of
edges which do not belong to any 1PI subgraph. Selecting any line from a graph
defines a {\em henge\/}, which is just the set of 1PI components of the graph
with the specified line removed. An example of a henge is $\scriptstyle
\diagchar{'102}$, where the heavy lines indicate the set of 1PI subgraphs in the
henge corresponding to the light line. We shall write $\G/\H$ to indicate the
graph obtained by shrinking each 1PI subgraph $\Th$ in $\H$ to a point. If $\G$
is a 1PI graph and $\L\in\G$ some edge, then $\G$ may be considered as a single
loop $\G/\H(\G,\L)$ with the 1PI subgraphs in the henge $\H(\G,\L)$ acting as
``effective vertices.'' For the example above the graph $\G/\H$ is
$\scriptstyle\diagchar{'14}$. The set of all henges for a four-loop contribution
to the two-point function of $\phi^3$ theory is $\Bigl\{\diagop,from'75to'104.
\Bigr\}$; the henges $\H(\G,\L)$ shown as heavy lines correspond to $\L$ being
any of the light lines.

We define $I_\lambda(\G)$ to be the Feynman integral corresponding to $\G$ where
all the lines carry momentum greater than $\lambda$; that is $|k_\L| >
\lambda\quad(\forall\L\in\G)$ where we use the usual Euclidean norm. This
corresponds to Feynman rules in which an extra step function $\theta(|k_\L|^2 -
\lambda^2)$ is associated with each line. $i_\lambda(\G)$ is defined to be the
integrand of the graph $\G$.

\section{The $\Rop$ operation}

We now apply the simple momentum space decomposition which says that at every
point in the space of loop momenta $k$ some line $\L$ has to carry the smallest
momentum:
\begin{displaymath}
  I_\lambda(\G) = \sum_{\L\in\G} \int_\lambda^\infty dk\,
    i_k\bigl(\G/\H(\G,\L)\bigr) \prod_{\Th\in\H(\G,\L)} I_k(\Th).
\end{displaymath}
For each henge all possible subdivergences of $I(\G)$ must live within one of
the ``effective vertices''~$\Th$, so it is most natural to define the $\Rbar$
operation, which subtracts all subdivergences, as
\begin{equation}
  \Rbar I_\lambda(\G) \equiv \sum_{\L\in\G} \int_\lambda^\infty dk\,
    i_k\bigl(\G/\H(\G,\L)\bigr) \prod_{\Th\in\H(\G,\L)} \Rop I_k(\Th), \qquad
  \Rop I_\lambda(\G) \equiv \Rbar I_\lambda(\G) - \K\Rbar I_0(\G),
  \label{eq:r-def}
\end{equation}
where $\Rop$ is the operation which subtracts all divergences.

\subsection{The subtraction operation $-\K$}

The subtraction operator $-\K$ removes the divergent part of $I(\G)$. Various
choices are possible
\begin{itemize}
\item For minimal subtraction $-\K I(\G)$ subtracts the pole terms in the
  Laurent expansion of $I(\G)$ in the dimension $D$. In this case the BPH
  theorem states that these subtractions are local; i.e., polynomial in the
  external momenta $p$.
\item $\K$ can be chosen to be the Taylor series subtraction operator
  $T^{\deg\G}I(\G)$ with respect to the external momenta $p$, where $\deg\G$ is
  the overall (power counting) degree of divergence of $\G$. In this case the
  BPH theorem states that the subtracted Feynman integrals are convergent, i.e.,
  they have a finite limit as the cutoff $\Lambda\to\infty$.
\end{itemize}
The subtraction operation commutes with differentiation: for minimal subtraction
$[\partial,\K] = 0$ trivially, whereas for Taylor series subtraction $\partial
T^n = T^{n-1}\partial$ but, as we shall show in
sections~\ref{sec:bounding-inequalities} and~\ref{sec:bph-proof},
$\deg\partial\G = \deg\G - 1$, so $[\partial,T^{\deg}] = 0$.

Strictly speaking we define $-\K$ to replace the divergent part with a finite
polynomial of degree $\deg\G$ in the external momenta. The finite part of a
subtracted graph is specified unambiguously by some set of {\em renormalization
conditions\/}, which fix the values of $I(p_0), \partial
I(p_0),\ldots,\partial^{\deg\G}I(p_0)$ at the {\em subtraction point\/} $p_0$.
As these renormalization conditions have no loop corrections this only affects
tree-level diagrams, and leads us to the following conventions\footnote{These
are different from the usual conventions, but make the formalism tidier.} for
the graph $v$ consisting of a single vertex: $\Rop I_\lambda(v) = I_0(v)$,
$\Rbar I_\lambda(v) = 0$, and $-K\Rbar I_0(v) = I_0(v)$.

\subsection{Equivalence to Bogoliubov's Definition}

A {\em spinney\/} \cite{kennedy82b} is a covering of a graph by a set of
disjoint 1PI subgraphs. Single vertices are allowed as elements of
spinneys:\footnote{This differs from the usual definition because of our
conventions for single vertices.} in other words, all the vertices of a graph
are included in a spinney, but not necessarily all the edges. The {\em wood\/}
$\W(\G)$ is the set of all spinneys for a graph $\G$. Note that every henge is a
spinney, but not vice versa. We shall use the notation $I_\lambda\left(
\G/\S\star\prod_{\Th\in\S}f(\Th) \right)$ to mean the Feynman integral for the
graph $\G/\S$ where all internal lines carry momentum larger than $\lambda$ and
the function $f(\Th)$ is the Feynman rule for the ``effective vertex'' $\Th$.
The {\em proper wood\/} $\bar\W(\G)$ is just the wood with the spinney $\S=\G$
omitted. The following is an example of a wood from $\phi^3$ theory:
$\W\Bigl(\diagchar{'36} \Bigr) = \Bigl\{\diagop,from'36to'105.\Bigr\}$.

Bogoliubov's \cite{bogoliubov56a,bogoliubov57a} definition of the $\Rop$
operation is
\begin{eqnarray}
  \Rbar_BI_\lambda(\G) &\equiv& \sum_{\S\in\bar\W(\G)} I_\lambda\left(\G/\S
    \star\prod_{\D\in\S} -K\Rbar_BI_0(\D)\right), \nonumber \\
  \Rop_BI_\lambda(\G) &\equiv& \Rbar_BI_\lambda(\G) - K\Rbar_BI_0(\G)
     = \sum_{\S\in\W(\G)}I_\lambda\left(\G/\S
       \star\prod_{\D\in\S}-K\Rbar_BI_0(\D)\right),
  \label{eq:bogoliubov}
\end{eqnarray}
where we have made a generalization to allow a non-vanishing~$\lambda$. We shall
prove the equivalence of our definition of equation~(\ref{eq:r-def}) with that
of equation~(\ref{eq:bogoliubov}) by induction. For graphs with no loops the
equivalence is trivial, and we assume that for graphs with fewer than $L$ loops
$\Rop I_\lambda(\D)=\Rop_BI_\lambda(\D)$. For an $L$ loop graph $\G$
equation~(\ref{eq:r-def}) gives us
\begin{eqnarray}
  \Rbar I_\lambda(\G) &=& \sum_{\L\in\G} \int_\lambda^\infty dk\,
    i_k(\G/\H) \prod_{\Th\in\H(\G,\L)} \sum_{\S\in\W(\Th)} I_k\left(\Th/\S
      \star\prod_{\D\in\S} -K\Rbar_BI_0(\D)\right) \nonumber \\
  &=& \sum_{\L\in\G} \int_\lambda^\infty dk\,
    i_k(\G/\H) \prod_{\S\in\W\bigl(\H(\G,\L)\bigr)} I_k\left(\H/\S
      \star\prod_{\D\in\S} -K\Rbar_BI_0(\D)\right),
    \label{eq:*}
\end{eqnarray}
where we have defined $\W(\H)$ to be the set of all spinneys which lie within
$\H$.\footnote{To be precise, $\W(\H)$ is the set of spinneys which are sets of
disjoint 1PI subgraphs which are either single vertices or are subgraphs of one
of the 1PI subgraphs of $\G$ in the henge $\H$.} It is clear that
$\S\in\bar\W(\G)\implies\exists\L:\S\in\W\bigl(\H(\G,\L)\bigr)$: just choose any
line $\L\not\in\S$. It is also clear that $\S\in\W\bigl(\H(\G,\L)\bigr)\implies
\S\in\bar\W(\G)$: just note that $\L\not\in\H(\G,\L)$. Hence all of the
subtractions in equation~(\ref{eq:*}) correspond exactly to the spinneys in
$\bar\W(\G)$, and the subtractions corresponding to some such spinney are
\begin{displaymath}
  \sum_{\L\in\G:\S\in\W\bigl(\H(\G,\L)\bigr)} \int_\lambda^\infty dk\,
    i_k(\G/\H) I_k\left(\H/\S \star\prod_{\D\in\S} -K\Rbar_BI_0(\D)\right)
  = I_\lambda\left(\G/\S \star\prod_{\D\in\S} -K\Rbar_BI_0(\D)\right),
\end{displaymath}
because the set of lines $\L$ for which $\S\in\W\bigl(\H(\G,\L)\bigr)$ is
precisely $\G/\S$. Therefore we have shown that $\Rbar I_\lambda(\G) =
\Rbar_BI_\lambda(\G)$.

The definition of the $\Rop$-operation can be made even more explicit and less
recursive using Zimmermann's \cite{zimmermann70a} forest notation: however it is
easier to construct proofs and write computer programs to automate
renormalization using recursive definitions.

In Bogoliubov's form it is manifest that $[\partial,\Rop]=0$, because
(i)~$[\partial,\K]=0$, and (ii)~the definition of $\Rop$ is purely graphical,
and the graphical structure is not changed by differentiation.

\subsection{Equivalence to Counterterms}

We shall show that the subtractions made by the $\Rop$ operation are equivalent
to the addition of counterterms to the action. As this is a purely combinatorial
proof it is convenient to use the generating functional $Z(J) = \langle
e^{-S(\phi) + J\phi} \rangle = \exp\left[ -S_I\left( \textstyle\ddJ\right)
\right] e^{\half J\Delta J}$, where $S(\phi) = \half\phi\Delta^{-1}\phi +
S_I(\phi)$. Perturbation theory may be viewed as an expansion in the number of
vertices in a graph,
\begin{displaymath}
  Z(J) = \sum_{n=0}^\infty {(-)^n\over n!}
    \left[S_I\left(\textstyle\ddJ\right)\right]^n e^{\half J\Delta J}
  = \sum_{n=0}^\infty {(-)^n\over n!} \sum_{\G_n}I\bigl(\G_n(J)\bigr)
    e^{\half J\Delta J};
\end{displaymath}
where the last sum is over all graphs $\G_n$ containing exactly $n$
vertices and which have $J$ attached to their external legs. We define the
{\em renormalized generating functional\/} as
\begin{displaymath}
  \Rop Z(J) \equiv \sum_{n=0}^\infty {(-)^n\over n!} \sum_{\G_n} \Rop I(\G_n)
    e^{\half J\Delta J}
  = \sum_{n=0}^\infty {(-)^n\over n!}
    \sum_{\G_n} \sum_{\S\in\W(\G_n)}
      I\biggl(\G/\S\star\prod_{\D\in\S}-\K\Rbar I(\D)\biggr)
	e^{\half J\Delta J}.
\end{displaymath}
Using the identity
\begin{displaymath}
  \sum_{\G_n} \sum_{\S\in\W(\G_n)}\prod_{\D\in\S}-\K\Rbar I(\D)
    = \!\!\sum_{\stackrel{r_0,\ldots,r_n}{r_0 + \cdots + r_n = n}}
      \!\!{n!\over\prod_{j=0}^n j!^{r_j} r_j!}
	\prod_{j=0}^n \biggl[\sum_{\G_j} -\K\Rbar I(\G_j)\biggr]^{r_j}
\end{displaymath}
where the last sum is over all graphs $\G_j$ with exactly $j$ vertices, we
obtain
\begin{eqnarray*}
  RZ(J) &=& \sum_{n=0}^\infty {(-)^n\over n!}
    \!\!\sum_{\stackrel{r_0,\ldots,r_n}{r_0 + \cdots + r_n = n}}
      \!\!{n!\over\prod_{j=0}^n j!^{r_j} r_j!}
	\prod_{j=0}^n \biggl[\sum_{\G_j} -\K\Rbar
	  I\bigl(\G_j({\textstyle\ddJ})\bigr)\biggr]^{r_j}
	    e^{\half J\Delta J} \\ 
  &=& \prod_{j=0}^\infty \sum_{r_j=0}^\infty {1\over r_j!}
    \biggl[{1\over j!} \sum_{\G_j} \K\Rbar
      I\bigl(\G_j({\textstyle\ddJ})\bigr)\biggr]^{r_j} e^{\half J\Delta J}
  = \prod_{j=0}^\infty \exp\biggl[{1\over j!} \sum_{\G_j} \K\Rbar
    I\bigl(\G_j({\textstyle\ddJ})\bigr)\biggr] e^{\half J\Delta J}\\
  &=& \exp\left[\K\Rbar e^{S_I\left(\ddJ\right)}\right]
    e^{\half J\Delta J}.
\end{eqnarray*}
We have thus shown that $RZ(J) = \langle e^{-S_B(\phi) + J\phi}\rangle$,
with the {\em bare action\/}
\begin{equation}
  S_B(\phi) = \half\phi\Delta^{-1}\phi - \K\Rbar e^{S_I(\phi)}.
  \label{eq:bare-action}
\end{equation}

Observe that there is no simple one to one correspondence between countergraphs
and subtractions, but that the combinatorial factors arrange themselves
correctly. The counterterms are monomials in the bare action, and we draw the
fields $\phi$ or functional derivatives $\ddJ$ by open circles at the end of the
amputated external legs; such graphs are symmetric under interchange of their
external legs. The appropriate combinatorial factors must be used for each
graph. Some of the counterterms in $\phi^3$ theory in $D$ dimensions are
\begin{displaymath}
  \begin{array}{rcl@{\qquad}rcl}
    \diagchar{'20} &=& {1\over2} \diagchar{'26} + {1\over4} \diagchar{'33}
	+ O(\hbar^3) &
      \diagchar{'22} &=& \diagchar{'30} + {3\over2} \diagchar{'35}
	+ {3\over4} \diagchar{'13} + O(\hbar^3) \\
    \diagchar{'21} &=& {1\over2} \diagchar{'27} + {1\over2} \diagchar{'34}
	+ \diagchar{'12} + O(\hbar^3) &
      \diagchar{'23} &=& 3 \diagchar{'31} + O(\hbar^2)
  \end{array}
\end{displaymath}
The countergraphs built using these counterterms correspond to subtractions in
the following way:
\begin{displaymath}
  \begin{array}{c||c|c|c|c|c}
    & {1\over2} \diagchar{'14} & {1\over2} \diagchar{'16} &
      {1\over2} \diagchar{'15} & \diagchar{'17} & \diagchar{'25} \\
    \hline\hline
    {1\over2} \diagchar0 & {1\over2} \diagchar1 & {1\over2} \diagchar2 &
    {1\over2} \diagchar3 && {1\over2} \diagchar4 \\
    \hline
    {1\over2} \diagchar5 & {1\over2} \diagchar7 + {1\over2} \diagchar{'10} &&&
      {1\over2} \diagchar{'6} & {1\over2} \diagchar{'11}
  \end{array}
\end{displaymath}

\section{Bounding Inequalities}
\label{sec:bounding-inequalities}

A condition for applicability of our proof of the BPH theorem is that we require
certain bounds to hold at tree level. All vertices and propagators $\D$ satisfy
$|I_\lambda(\D)| \leq \const\chi(\lambda)^{\deg\D}$, where $c$ is a constant,
and the {\em overall degree of divergence\/} $\deg\D$ is a number which will be
used for power counting. The monotonically increasing bounding function $\chi$
must satisfy
\begin{equation}
  \int_\lambda^\infty dk\, \chi(k)^\nu \leq \const\chi(\lambda)^{\nu+1}
    \quad (\nu+1<0), \qquad
  \int_0^\lambda dk\, \chi(k)^\nu \leq \const\chi(\lambda)^{\nu+1+0}
    \quad (\nu+1\geq0).
  \label{eq:bound-integrals}
\end{equation}
Differentiation with respect to external momenta must lower the degree of
divergence, $\deg(\partial\G) = \deg\G-1$. This means that we also require that
all derivatives of vertices and propagators must satisfy the bounds
\begin{equation}
  |\partial^nI_\lambda(\D)| \leq \const\chi(\lambda)^{\deg\D-n}.
\label{eq:tree-level-bound}
\end{equation}
The simple choice $\chi(k) \equiv \max(m,k)$ suffices for our proof of the BPH
theorem: a simple generalization is needed to prove the decoupling theorem. For
the lattice propagator $\Delta(k) = \left[m^2 + {4\over a^2} \sum_{\mu=1}^D
\sin^2\half k_\mu a\right]^{-1}$ the inequality $2|x|/\pi \leq |\sin x| \leq
1$ may be used to show that (\ref{eq:tree-level-bound}) holds.

For ``sharp cutoffs'' like a lattice regulator for which each propagator
has a factor $\theta(k_\L - \pi/a)$ there are also ``surface terms'' which
arise in derivatives of graphs. The generalization of
equation~(\ref{eq:tree-level-bound}) to include these terms is straightforward.

\section{Proof of the BPH Theorem}
\label{sec:bph-proof}

Our proof uses the induction hypothesis that $|\Rop I_\lambda(\G)| \leq \const
\chi(\lambda)^{\deg\G+0}$ for all graphs with fewer than $L$ loops. For $L=0$
this follows trivially from the  bounding inequalities of the previous section.
To show that it continues to hold for $L$-loop graphs we consider two cases.

\subsection{Overall convergent diagrams with $L$ loops}

From the definition (\ref{eq:r-def}) of $\Rbar$ we have
\begin{displaymath}
  |\Rbar I_\lambda(\G)| \leq \sum_{\L\in\G} \int_\lambda^\infty dk\,
    |i_k(\G/\H)| \prod_{\Th\in\H(\G,\L)} |\Rop I_k(\Th)|.
\end{displaymath}
Using the induction hypothesis for the subgraphs $\Th$ and the tree level
bounds~(\ref{eq:tree-level-bound}) for $i_k(\G/\H)$ we get
\begin{displaymath}
  |\Rbar I_\lambda(\G)| \leq \const \sum_{\L\in\G}
    \int_\lambda^\infty dk\,\chi(k)^{\deg(\G/\H)-1}
      \prod_{\Th\in\H(\G,\L)} \chi(k)^{\deg\Th+0}
  = \const \sum_{\L\in\G} \int_\lambda^\infty dk\,\chi(k)^{\deg\G-1+0};
\end{displaymath}
and upon integrating the bounding function using~(\ref{eq:bound-integrals}) we
find $|\Rbar I_\lambda(\G)| \leq \const \chi(\lambda)^{\deg\G+0}$ for
$\deg\G<0$. This establishes the induction hypothesis, since $\Rop I_\lambda(\G)
= \Rbar I_\lambda(\G)$ in this case.

\subsection{Overall divergent diagrams with $L$ loops}

Taylor's theorem for the function $\Rbar I_0(p)$ gives
\begin{displaymath}
  \Rbar I_0(p) = T^{\deg\G}\Rbar I_0(p) +
    \int_{p_0}^p dp_1 \ldots \int_{p_0}^{p_{\deg\G}} dp_{\deg\G+1}\,
      \partial^{\deg\G+1}\Rbar I_0\left(p_{\deg\G+1}\right),
\end{displaymath}
and since $\Rbar$ and $\partial$ commute
\begin{displaymath}
  \Rbar I_0(p) = T^{\deg\G}\Rbar I_0(p) +
    \int_{p_0}^p dp_1 \ldots \int_{p_0}^{p_{\deg\G}} dp_{\deg\G+1}\,
      \Rbar\partial^{\deg\G+1}I_0\left(p_{\deg\G+1}\right).
\end{displaymath}
The (sum of) graphs $\partial^{\deg\G+1}\G$ are overall convergent, so the
integrand is finite, and as the integral is over a compact region any
divergences must be in the polynomial part. The $\Rop$ operation removes this
polynomial $T^{\deg\G}\Rbar I_0(p)$, so $|\Rop I_0(p)| \leq \int_{p_0}^p dp_1
\ldots \int_{p_0}^{p_{\deg\G}} dp_{\deg\G+1}\, \left|R\partial^{\deg\G+1}
I_0\left( p_{\deg\G+1} \right)\right|$. Using the inductive bound already
established for the overall convergent integrand $|\Rop I_0(p)| \leq
\int_{p_0}^p dp_1 \ldots \int_{p_0}^{p_{\deg\G}} dp_{\deg\G+1}\, \const
\chi(0)^{-1+0} \leq \const \chi(0)^{\deg\G+0}$ we prove that $\Rop I_0(\G)$ is
made finite by local subtractions, but we still need to establish the induction
hypothesis. In the definition of $\Rbar I_0(\G)$ we may split the integration
region $\int_0^\infty dk=\int_0^\lambda dk+\int_\lambda^\infty dk$, hence
\begin{displaymath}
  \Rop I_0(\G) = \Rop I_\lambda(\G) +
    \sum_{\L\in\G} \int_0^\lambda dk\,
      i_k(\G/\H) \prod_{\Th\in\H(\G,\L)} \Rop I_k(\Th).
\end{displaymath}
All that is left to do is to bound the integral over the ``infrared region''
using essentially the same technique as for the overall convergent case above
\begin{eqnarray*}
  |\Rop I_\lambda(\G)| &\leq& \const \chi(0)^{\deg\G+0}
      + \const \sum_{\L\in\G} \int_0^\lambda dk\,\chi(k)^{\deg(\G/\H)-1}
      \prod_{\Th\in\H(\G,\L)} \chi(k)^{\deg\Th+0} \\
    &\leq& \const \chi(0)^{\deg\G+0} + \const \sum_{\L\in\G}
      \int_0^\lambda dk\,\chi(k)^{\deg\G-1+0}
	\leq \const \chi(\lambda)^{\deg\G+0} \qquad (\deg\G\geq0).
\end{eqnarray*}

\section{Power Counting}

Consider a connected Feynman diagram $\G$ in a $D$ dimensional field theory with
an arbitrary polynomial action. Let it have $I_a$ lines of type $a$, $V_b$
vertices of type $b$, and $E_a$ external legs of type $a$. Let $n_{ab}$ be the
number of lines of type $a$ which are attached to vertex $b$, $d'_b$ be the
degree of this vertex, and $d_a$ be the degree of lines of type~$a$. Every line
has to end on an appropriate vertex, so $\sum_{b} n_{ab}V_b = E_a + 2I_a\;
(\forall a)$. We require exactly $V-1$ lines to connect $V$ vertices into a
tree; every extra line produces a loop: hence $L = I - V + 1 = \sum_{a}I_a -
\sum_{b}V_b + 1$. The overall degree of the graph can be obtained by counting,
$\deg\G = LD + \sum_{b}V_bd'_b + \sum_{a}I_ad_a$. Eliminating $L$ and $I_a$ from
these equations we obtain
\begin{displaymath}
  \deg\G = \sum_{b} V_b \biggl[ \half\sum_{a}
    \left\{n_{ab} (d_a + D)\right\} + d'_b - D \biggr]
    - \half\sum_{a} E_a(d_a + D) + D.
\end{displaymath}
The {\em dimension\/} of the monomial $\V_b$ in the action\footnote{This differs
by $D$ from the dimension of the corresponding monomial in the Lagrangian
density.} corresponding to the vertex of type $b$ may be defined to be $\dim\V_b
\equiv \sum_{a} n_{ab}\dim(\phi_a) + d'_b - D$, where the dimension of the field
$\phi_a$ is defined such that the dimension of its kinetic term in the action
vanishes; that is, $\dim\phi_a \equiv \half(d_a + D)$. This gives $\dim\V_b =
\half\sum_{a} n_{ab}(d_a + D) + d'_b - D$. We thus obtain
\begin{equation}
  \deg\G = \sum_{b} V_b \dim\V_b - \sum_{a} E_a \dim\phi_a + D.
  \label{eq:power-counting}
\end{equation}

A theory is {\em superrenormalizable}, that is has only a finite number of
overall divergent graphs, if the coefficients of $V_b$ are negative: $\dim\V_b <
0\;(\forall b)$. The theory is {\em renormalizable}, that is only a finite
number of Green's functions are overall divergent, if none of the coefficients
of $V_b$ are positive, $\dim\V_b\leq 0\;(\forall b$), and all the coefficients
of $E_a$ are positive, $\dim\phi_a>0\quad(\forall a)$. In general, all local
monomials of dimension $\leq0$ will be required as counterterms.

\section{Some Applications}

\subsection{Operator Insertions}

Let $\Omega(\phi)$ be an operator which is local and polynomial in the field
$\phi$. Add a source term for $\Omega$ into the action, $Z(J,J') \equiv \langle
e^{-S(\phi) + J\phi + J'\Omega(\phi)}\rangle$. The BPH theorem tells us that
this theory can be renormalized by adding local counterterms of the form given
by equation~(\ref{eq:bare-action}), $\Delta S(\phi,J') = - S_I(\phi) +
J'\Omega(\phi) -\K\Rbar\exp\left[S_I(\phi) - J'\Omega(\phi)\right]$.  Expanding
in powers of $J'$ gives $\Delta S(\phi,0) + J'\K\Rbar\left[e^{S_I(\phi)}
\Omega(\phi)\right] + J'\Omega(\phi) + O({J'}^2)$. We may associate the
counterterms linear in $J'$ with the operator to define a renormalized operator
$N(\Omega) \equiv -\K\Rbar\left[e^{S_I(\phi)} \Omega(\phi)\right]$. Power
counting (\ref{eq:power-counting}) tells us that $\deg\G = V_\Omega\dim\Omega +
\sum_{b} V_b\dim\V_b - \sum_{a} E_a \dim\phi_a + D$, where $V_\Omega$ are the
number of $\Omega$ vertices in $\G$. As we are interested in a single insertion
of $\Omega$ we only care about counterterms linear in $J'$, and these get
contributions only from diagrams $\G$ with $V_\Omega=1$. Thus $\deg\G \leq
\dim\Omega + D$, for a renormalizable theory, which means that we only get
counterterms of dimension $\leq\dim\Omega$. Analogous arguments easily establish
the operator product expansion.

\subsection{Oversubtraction}

If one subtracts more than $\deg\G+1$ terms in the Taylor expansion in the
external momenta from a graph then the graph does not become any more convergent
(in the sense of lowering the exponent in the inductive bound of
section~\ref{sec:bph-proof}), but the dependence on the cutoff parameter is
reduced. This result of Zimmermann's \cite{zimmermann70a} is central to
Symanzik's improvement programme on the lattice \cite{symanzik83a}. Our methods
may be used to prove Zimmermann's theorem by applying the arguments of
section~\ref{sec:bph-proof} to the derivatives of $I(\G)$ with respect to the
cutoff $a=1/\Lambda$.

\section*{Acknowledgements}
I would like to thank Bill Caswell for his collaboration on the early stages of
this work, Joe Sucher for encouragement over the (many) years these ideas have
been germinating, and Urs Heller for discussions of renormalization on the
lattice. I would also like to thank the editors of these proceedings for their
forbearance in allowing me to to exceed their length recommendations.

This research was supported by by the U.S. Department of Energy through Contract
Nos. DE-FG05-92ER40742 and DE-FC05-85ER250000.

\bibliographystyle{utphys}
\bibliography{adk,renormalization}

\end{document}